\documentclass[iop]{emulateapj}

\begin{document}

\title{Microphysics of Neutron Star Outer Envelopes in the Periodized, Magnetic Thomas-Fermi Model}

\shorttitle{Microphysics of Neutron Star Outer Envelopes} 

\author{T. A. Engstrom$^1$, V. H. Crespi$^1$, B. J. Owen$^{2,3}$, J. Brannick$^4$, and Xiaozhe Hu$^5$}
\affil{$^1$Department of Physics, The Pennsylvania State University, University Park, PA 16802, USA}
\affil{$^2$Department of Physics, Institute for Gravitation and the Cosmos, Center for Particle and Gravitational
Astrophysics, \\The Pennsylvania State University, University Park, PA 16802, USA}
\affil{$^3$Department of Physics, Texas Tech University, Lubbock, TX 79409, USA}
\affil{$^4$Department of Mathematics, The Pennsylvania State University, University Park, PA 16802, USA}
\affil{$^5$Department of Mathematics, Tufts University, Medford, MA 02155, USA}
\email{tae146@psu.edu\\vhc2@psu.edu\\bjo10@psu.edu\\brannick@psu.edu\\xiaozhe.hu@tufts.edu}

\begin{abstract}
Static and dynamic properties of low density outer envelopes of neutron stars are calculated within the nonlinear magnetic Thomas-Fermi model, assuming degenerate electrons.  A novel domain decomposition enables proper description of lattice symmetry and may be seen as a prototype for the general class of problems involving nonlinear charge screening of periodic, quasi-low-dimensionality structures, e.g.\ liquid crystals.  We describe a scalable implementation of the method using \textit{Hypre}.  Phase velocity of long wavelength transverse phonons is found to be a factor of 5-7 larger than in the corresponding Coulomb crystal model, which could have implications for low temperature phonon-mediated thermal conductivity.  Other findings include $c'<0$ elastic instabilities for both bcc and fcc lattices, reminiscent of the situation in some light actinides, and suggestive of a symmetry-lowering transition to a tetragonal or orthorhombic lattice.
\end{abstract}


\notetoeditor{Regarding Equations 7-12 (without mathletters) or Equations 7a-8c (with mathletters):  Ideally there would be no page break between these equations,  and the "=" signs in all six equations would be vertically aligned to emphasize that they are part of the same problem.}

\section{Introduction}

In conventional solid state physics, the Thomas-Fermi model is regarded as an historical development and a pedagogical tool (although it remains a key ingredient of modern, orbital-free density functional theory). Conversely, in certain extreme conditions of solid state astrophysics, where order-of-magnitude estimates of thermodynamic quantities are sought, the failure to predict binding/condensation of atoms is not a serious deficiency due to matter being under high pressure, and appropriate \textit{ab initio} methods are in a state of infancy, the Thomas-Fermi model has not yet faded into obsolescence.  In particular, \textit{magnetic} Thomas-Fermi models first written down in the 1970s, and extended in many directions in the 1980-90s, continue to be relied upon for the equation of state of magnetized neutron star outer envelopes -- see \citet[chap. 4]{hae07}, the review by \citet{lai01}, and references within.  Magnetic Thomas-Fermi (MTF) models are aimed at matter composed of heavy atoms in a highly non-perturbative magnetic field, and are reasonably appropriate for the outer $\sim10$ meters (in this paper, ``outer envelopes") of many neutron stars, where $\rho <10^6$ g/cc and $B\sim10^{12}$-$10^{13}$ gauss.  For a free electron, $10^{13}$ gauss corresponds to a magnetic length $65$ times smaller than the Bohr radius $a_0$, and zero-point cyclotron energy (or Zeeman energy) more than $1/10$ of the rest mass.  This physical regime has also been treated by Kohn-Sham density functional theory \citep{med06,jon86} -- potentially much more accurate than the MTF model, which is only asymptotically exact.  However, these more sophisticated calculations have not yet been made fully self-consistent (in the case of 3D condensed matter), and the focus has been on prediction of binding energies and the related question of magnetic condensation at neutron star surfaces \citep{med07, pot13}, rather than lattice dynamical properties which are the main focus of this work. Lattice dynamical properties have been extensively studied in a higher density, completely pressure ionized regime, using Coulomb crystal models (see \citet{bai12}, \citet{bai13}, and included references). 

An under-appreciated property of the MTF model is that its regime of asymptotic exactness contains many field configurations for which Bloch's theorem can be proved.  (In other huge magnetic field regimes, use of periodic boundary conditions may not be so innocuous.)  But as with its nonmagnetic counterpart, it has been standard practice to replace the properly periodized MTF model with an approximate version having spherical Wigner-Seitz cells and a vanishing normal derivative at the cell edge.  Physics related to bulk phase stability is out of the reach of this approximation, due to the lack of an explicit lattice. Here we attempt to extract some of this physics by extending a novel domain decomposition approach invented by \citet{mac83}.  While M\&H could not reach the threshold accuracy required to resolve subtle energy differences between structures, we significantly improve their method by incorporating curved subdomain interfaces with the appropriate symmetry, and implement the improved method making use of the \textit{Hypre} library of scalable, multigrid-preconditioned solvers \citep{fal06}.  We then calculate the equation of state, phase diagram, elastic constants, Brillouin zone edge phonon frequency, and exchange correction in the periodized MTF model.  Some of these calculations were run on the Stampede cluster through XSEDE \citep{tow14}. 

To the extent that calculations involve energy \textit{differences} between structures, states of strain, etc., the divergence of the electron density at the nuclei predicted by the MTF model is not necessarily a significant source of error. The incorrectly described regions, close to the nuclei, have a weak dependence on lattice structure, and thus their contribution to relative phase stabilities or elastic response tends to cancel out. Similar fortuitous cancellations are seen ubiquitously in lower pressure electronic structure calculations at the density functional level \citep{cap06}, and underlie the enormous success of the local density approximation in predicting structural phase diagrams, phonon frequencies, and elastic constants (see for example, \citet{he14}).  Other properties such as formation energies of solids from atomic constituents -- where one of the comparison systems does not form a compact lattice -- do require more careful treatment of gradients. The starting point for improving upon a Thomas-Fermi model is the inclusion of a Weizs{\"a}cker (gradient) term in the kinetic energy functional.  Unfortunately, the form of the Weizs{\"a}cker correction to the MTF model is not generally known \citep{lai01}, although it has been explored in certain limits \citep{fus92}. Here we confine ourselves to physical quantities that benefit from the cancellations described above.

Among our findings is a type of lattice instability that is typically driven by the splitting of sharp features in the electronic density of states due to a symmetry-lowering.  It is interesting that the MTF model, which has no ``density of states," should capture this kind of symmetry-lowering transition, and it is perhaps the simplest model that does so.

\section{Regime of Validity}

We begin by recapitulating the MTF regime as it pertains to an isolated, heavy atom \citep{fus92}.  Define the reduced field $b = B/B_0$ where $B_0 = m^2e^3c\hbar^{-3} = 2.4\times10^9$ gauss. A strong field is next defined as one in which the magnetic length $\ell = b^{-1/2}a_0$ beats out the zero-field mean electron spacing $Z^{-2/3}a_0$ as the smallest length scale in the problem: $b \gg Z^{4/3}$.  Instead of the standard three-dimensional behavior $k_F = (3\pi^2n_e)^{1/3}$, the Fermi momentum follows the lowest Landau level expression $k_{z,F} = 2\pi^2\ell^2n_e$ in a strong field $\mathbf{B} = B\hat{z}$, which modifies the usual $n_e^{5/3}$ dependence of the kinetic energy density to $n_e^3$.  MTF description additionally requires the electrostatic potential to vary slowly on the lengthscale $k_{z,F}^{-1}$, and simple scaling relations indicate this condition is met when $b \ll Z^3$.  MTF theory is an exact limit of quantum mechanics for $Z, bZ^{-4/3}\to\infty$ while $bZ^{-3}\to0$; in this regime, sphericity of the atom is not destroyed \citep{yng91,lie92}.

Extending MTF theory to bulk matter under pressure, one repeats the above arguments, replacing the characteristic size of an isolated atom with the Wigner-Seitz radius $r_s = (3Z/4\pi n_e^{\mathrm{avg}})^{1/3}$, and finds that the validity conditions put a restriction on the number of flux quanta penetrating the unit cell: $Z^{2/3} \ll (r_s/\ell)^2 \ll Z$.  

Like orbital-free density functional theory in general, the MTF model works with densities, not wavefunctions.  Wavefunctions are implicit in that the amplitude of the underlying wavefunction determines the density, but the phase is not directly expressed.  Therefore, the MTF model can't be expected to capture some implications of the phase structure of the underlying wavefunction, in particular, issues with periodicity-breaking in a strong magnetic field.  While this argument helps justify use of periodic boundary conditions (PBCs) for arbitrary field configurations in the MTF model, we can also show that PBCs are exact for a large number of field configurations satisfying $Z^{2/3} \ll (r_s/\ell)^2$.  Consequently, the physical regime treated by the MTF model here is also amenable to wavefunction-based methods that require (or are greatly simplified by) PBCs, and which may be used to check the accuracy of MTF predictions.  Consider the single-electron Hamiltonian
\begin{equation}
H = \frac{1}{2m}\Big(\mathbf{p}-\frac{e\mathbf{A}}{c}\Big)^2 + V(\mathbf{r}),
\end{equation}
where $V(\mathbf{r}) = V(\mathbf{r+R})$ and $\mathbf{R}$ is any lattice vector.  Magnetic translation operators $\mathcal{T}_R$ can be constructed that commute with $H$, but in general, they don't commute with each other -- the origin of Hofstadter periodicity breaking \citep{jai07, koh93}.  In the symmetric gauge
\begin{equation}
\mathcal{T}_R\mathcal{T}_{R'} = \mathcal{T}_{R'}\mathcal{T}_{R}\,\exp\bigg[ \frac{2\pi i}{\phi_0}\,\mathbf{B} \cdot (\mathbf{R} \times \mathbf{R}') \bigg], 
\end{equation}
where $\phi_0=hc/e$ is the flux quantum.  Now restrict the magnetic field orientation to be along an irreducible lattice vector $\mathbf{R}_{\parallel}$, where $R_{\parallel}\sim r_s$.  In other words, the field is oriented to a high symmetry direction in the crystal.  Choose $\mathbf{R}_{\parallel}$ as the primitive lattice vector $\mathbf{a}_1$.  Any valid choice of the remaining primitive vectors $\mathbf{a}_2$ and $\mathbf{a}_3$ gives flux $\phi=0$ through two elementary plaquettes (an elementary plaquette is defined by $\mathbf{a}_i\times\mathbf{a}_{j\neq i}$) and in the MTF regime, the flux through the third plaquette automatically satisfies $\phi/\phi_0\gg Z^{2/3}$.  Never is more than a small fractional adjustment of $B$ required for $\phi/\phi_0$ to be integer-valued through this third plaquette; with this adjustment $\{\mathcal{T}_{a_1},\mathcal{T}_{a_2},\mathcal{T}_{a_3},H\}$ form a commuting set and may be simultaneously diagonalized by the Bloch functions $\psi_{\mathbf{k},n}(\mathbf{r}) = e^{i\mathbf{k\cdot r}} u_{\mathbf{k},n}(\mathbf{r})$ where $n$ is a magnetic subband index and $|u_{\mathbf{k},n}|$ has the periodicity of the primitive lattice.  (For $\phi/\phi_0$ rational-valued through each elementary plaquette, $|u_{\mathbf{k},n}|$ is periodic over certain non-primitive cells; for irrational flux, $|u_{\mathbf{k},n}|$ is incommensurate with $V$).  Note that a change of basis takes the set of plaquette fluxes $\{0,0,\phi/\phi_0\in\mathbb{Z}\}$ into a different set of integer fluxes while the actual field configuration remains unchanged.

\section{Model \& Domain Decomposition}

The MTF model for a degenerate electron gas in the lowest Landau level interacting with a lattice of  point nuclei is defined by the energy functional
\begin{equation}
E[n_e] = E_{kin} + V_{ie} + V_{ee}, \label{functional}
\end{equation}
where
\begin{eqnarray}
E_{kin} & = & \frac{2\pi^4}{3b^2} \,\frac{e^2}{a_0}\, a_0^6 \int d^3r \, n_e^3(\mathbf{r}), \label{Ekin} \\
V_{ie} & = & -Ze^2 \sum_{\mathbf{R}} \int d^3r \frac{n_e(\mathbf{r})}{|\mathbf{r-R}|}, \label{Vie} \\
V_{ee} & = & \frac{e^2}{2} \int d^3r d^3r' \frac{n_e(\mathbf{r}) n_e(\mathbf{r}')}{|\mathbf{r-r'}|}. \label{Vee}
\end{eqnarray}
One obtains the MTF equation by imposing the stationarity condition $\delta \big( E - \mu\int d^3r\,n_e \big) = 0$ and combining the result with the Poisson equation $\Delta \Phi = 4\pi e n_e$.  While $\Phi$ is the total electrostatic potential from electrons and nuclei, nuclear charge density is omitted from the right hand side of this Poisson equation and instead taken into account via the boundary conditions $\Phi(\mathbf{r}) = Ze|\mathbf{r-R}|^{-1}$ as $|\mathbf{r-R}| \to 0$.  It is convenient to make a change of variables defined by $\mathbf{r} = \sigma\mathbf{x}$, $\mathbf{R} = \sigma\mathbf{X}$ and $\mu + e\Phi(\mathbf{r}) = Ze^2 u(\mathbf{x})/\sigma$, where $\sigma = a_0(Z\pi^2/8b^2)^{1/5}$.  Delaying further discussion of boundary conditions, we write down the startlingly concise PDE which is the result of these manipulations: $\Delta u = \sqrt u$.

\begin{figure}[t]
\plotone{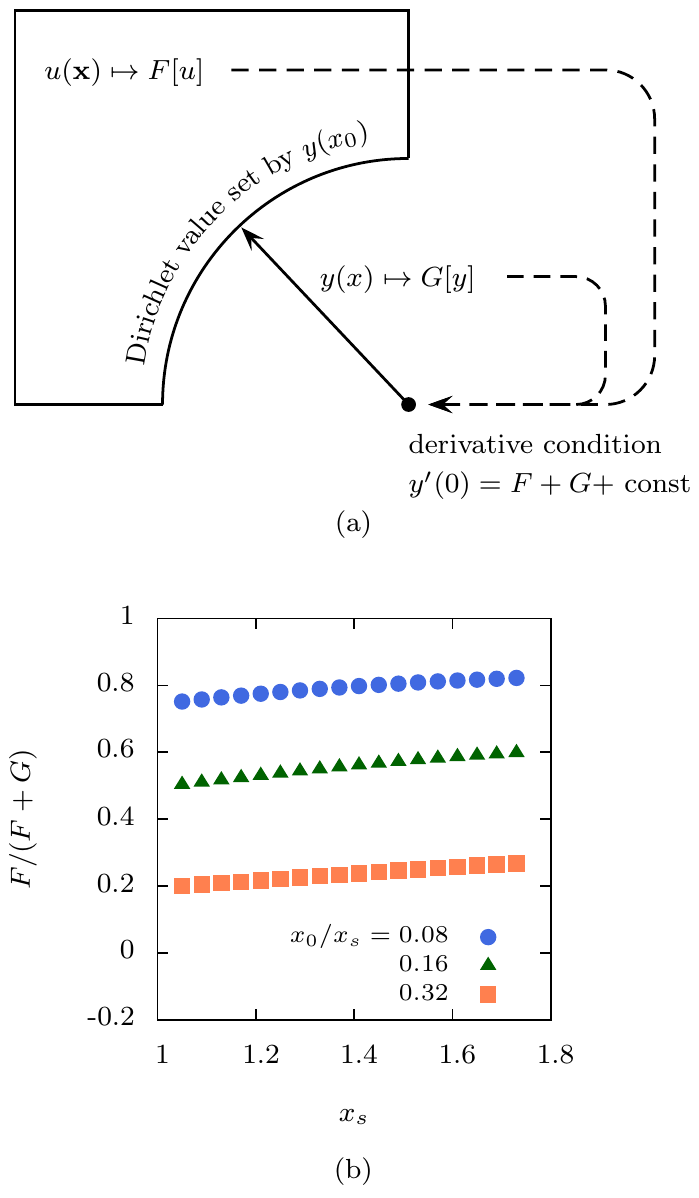}
\caption{(a) Schematic of information flow in the iMH domain decomposition.  (b) Fractional contribution of $F$ to the derivative condition $y'(0)$.  As indicated in the schematic, $F$ is the contribution from the large subdomain and $G$ is the contribution from all small subdomains. Data shown are from the last iteration (converged and normalized) of the bcc lattice problem.
\label{fig:flow}}
\end{figure}

To solve the model, we use an improved version of MacFarlane \& Hubbard's domain decomposition method, hereafter iMH.  The original method features subdomain interfaces that are easy to implement (boxes), but don't respect the symmetry of the solution, and we have observed that this method generates large and unphysical discontinuities in $\nabla u$ near the box corners.  Use of curved interfaces is the main improvement in iMH. Noting the method can be generalized to any lattice, we restrict the present discussion to cubic Bravais lattices, for which a convenient choice of domain $\Omega$ is one octant of the conventional unit cell.  Subdomain $\Omega_A$ is formed by centering a small sphere of radius $x_0$ on a lattice point, and taking the intersection with $\Omega$.   There are $N$ identical copies of $\Omega_A$, one in each corner of $\Omega$ where a nucleus is located (two corners for bcc, four for fcc).  A large ``swiss cheese" subdomain remains: $\Omega_B = \Omega \setminus (N\times\Omega_A)$.  A change of variable $y = xu$ in $\Omega_A$ removes the singularity at the nucleus and gives the boundary conditions $y(0)=1$ and $y'(0) = \sigma\mu/Ze^2 + \sigma\Phi(\mathbf{R})/Ze$, where $\Phi(\mathbf{R})$ is to be determined self-consistently.  (In the case of uniform background, this quantity is equal to the Madelung potential).  iMH is now written as a Schwarz alternating procedure, with first half-step given by the initial value problem
\begin{eqnarray}
y_k'' & = & \sqrt{x y_k}, \;\;\;\;\;\; 0\leq x \leq x_0, \label{ydoubleprime} \\
y_k(0) & = & 1, \\
y_k'(0) & = & \frac{2\xi}{x_s} + \frac{\epsilon_F}{Z}  -  \int_V d^3x \Big(\frac{\sqrt{u_{k-1}}}{4\pi} - \frac{1}{V}\Big) q(\mathbf{x}),\label{yprime}
\end{eqnarray} 
and second half-step given by the nonlinear boundary value problem
\begin{eqnarray}
\Delta u_k &=& \sqrt{u_k},\textrm{ in }\Omega_B, \label{pde} \\
\hat{n} \cdot \nabla u_k &=& 0,\textrm{ on flat parts of }\partial\Omega_B, \\
u_k &=& \frac{y_k(x_0)}{x_0},\textrm{ on curved parts of }\partial\Omega_B. \label{dirichlet}
\end{eqnarray}
In Equation \ref{yprime}, the $k^{th}$ derivative condition is obtained as a functional of the $k-1^{st}$ solution, extending the concept of ``overlap" in domain decompositions. Information flow in iMH is represented schematically in Figure \ref{fig:flow}(a). 
The integral over the primitive cell volume $V$ involves a product of the density nonuniformity correction and the Ewald-type sum   
\begin{eqnarray}
q(\mathbf{x}) & = & \frac{4\pi}{V}\sum_{\mathbf{G}\neq0}\frac{\cos(\mathbf{G}\cdot\mathbf{x}) \, e^{ -G^2/4\eta^2}}{G^2} \nonumber \\ & + & \sum_{\mathbf{X}}\frac{\mathrm{erfc}(\eta|\mathbf{X+x}|)}{|\mathbf{X+x}|} - \frac{\pi}{\eta^2 V} \label{latsum},
\end{eqnarray}
which can be obtained using a result due to \citet{nij57}.
The Madelung constant is given by $\xi = 2^{-1}x_s\lim_{\mathbf{x}\to0}(q(\mathbf{x}) - x^{-1}) = -0.8959293$ (bcc), $-0.8958736$ (fcc), while the Wigner-Seitz radius and Lagrange multiplier (Fermi energy) appear in dimensionless form $x_s=r_s/\sigma$ and $\epsilon_F = \mu\sigma/e^2$.

Decoupled from iMH is a normalization requirement $1 = (4\pi)^{-1}\int_V d^3x \sqrt u$ which completes the description of model. Since $\epsilon_F$ and $Z$ only appear together as a ratio, for a given lattice the model is specified by three parameters: $x_0$, $x_s$ and $\epsilon_F/Z$.  The Dirichlet radius $x_0$ has no physical significance, and we are interested in the limit $x_0 \to 0$, taken externally to the model ($x_0\to x_s$ essentially recovers the spherical Wigner-Seitz cell approximation).  Of the latter two parameters, only one is independent, as the other must be adjusted to its normalizing value. Once a converged, normalized solution is in hand, the $T=0$ Helmholtz free energy per nucleus is found by an integration over the primitive cell
\begin{equation}
U = \frac{Z^2e^2}{\sigma} \bigg[ \frac{y'(0)}{2} - \frac{1}{24\pi} \int_V d^3x\, u^{3/2} \bigg]. \label{helmholtz}
\end{equation}   
This expression excludes the energy stored in the magnetic field, which is assumed to be constant in the following calculations.

Comparison with linear response theory, which involves expanding the current model to leading order in $e\Phi(\mathbf{r})/\mu$, is facilitated by noting that $x_s$ in the current model is simply related to the inverse screening length $k_{TF}$ in the linear response theory
\begin{equation}
x_s = 1.43\,(k_{TF}r_s)^{2/5}. \label{lrt}
\end{equation}
For this reason we generally choose to fix $x_s$ and adjust $\epsilon_F/Z$ in the normalization procedure.  Finally, we give a formula for obtaining $x_s$ from conventional units
\begin{equation}
x_s = \frac{1.33\,b^{2/5}}{Z^{1/5}}\bigg(\frac{M \textrm{ in amu}}{\rho \textrm{ in g/cc}}\bigg)^{1/3}.
\end{equation}

\section{Numerical Implementation}

Starting with an initial guess $y_0'(0)$ corresponding to the uniform electron gas, the initial value problem (Equations \ref{ydoubleprime}--\ref{yprime}) is integrated using a semi-implicit Euler method with adaptive stepsize and built-in stability checks \citep{pre07}.  Having thus specified a Dirichlet value for the boundary value problem, we solve Equations \ref{pde}--\ref{dirichlet} by Newton's iteration with initial guess $u_0=1$.  Each linear system in Newton's iteration is solved by the finite volume method (FVM), through \textit{Hypre's} Struct interface, using either $600^3$ or $640^3$ gridpoints.  Certain 7-point stencils are modified to implement the Neumann and curved Dirichlet boundaries using standard discretizations that preserve both the overall $O(h^2)$ accuracy of the FVM scheme and the discrete maximum principle \citep{mor05}.  Each linear system is preconditioned by one V-cycle of \textit{Hypre}'s SMG multigrid and then iterated with conjugate gradients until the relative residual norm $<10^{-9}$.  This high tolerance is required for Newton's iteration to converge to the same solution regardless of whether weighted Jacobi or symmetric R/B Gauss-Seidel is used in the preconditioner.  Newton's iteration is terminated when the relative solution difference norm and relative nonlinear residual norm are both $<10^{-9}$. Integrations must then be performed to update the derivative condition $y_1'(0)$.  In $\Omega_B$, we apply a low-order quadrature rule cell-wise over dual FVM cells, where the approximate solution is trilinear-accurate.  Exceptions are the dual cells cut by a curved boundary; for $x_0/x_s = 0.08$ there are $\sim 10^4$ of these.  Each cut cell is handled by simple Monte Carlo integration with a few thousand points, which gives sufficient accuracy without causing a bottleneck.  We make use of \textit{Hypre's} internal ghost-zone updating routines for the dual cell-wise integration, and note this is a scalable approach.  A high-order quadrature rule is used for the $\Omega_A$ integration.  Iteration of iMH proceeds as described, the only difference being that convergence is accelerated by using the initial guess $u_k=u_{k-1}$ for the $k^{th}$ Newton's iteration.  As a check, we also try $u_k=1$ and find no dependence on which of these initial guesses is used.  iMH is iterated until $||u_k - u_{k-1}|| / || u_k|| < 10^{-9}$. 

While the limit $x_0 \to 0$ is desired from a physical standpoint, it must be kept in mind that the Dirichlet boundaries can be represented as smoothly curved surfaces within the structured grid only for $x_0/h \gg 1$.  Increasing $x_0$ increases the rate of information transfer from $\Omega_A$ to $\Omega_B$, but it also decreases the rate of information transfer in the reverse direction, illustrated by Figure \ref{fig:flow}(b).  We therefore restrict our study of $x_0$-dependence to the compromise range where $x_0/x_s$ goes from 0.08 to 0.32. Over this range, one can discern no discontinuity in $\nabla u$ at the interface, as occurs in the original MacFarlane \& Hubbard method, and the SMG-preconditioned solver yields consistent results.  \textit{Hypre's} PFMG preconditioner also yields consistent results for the larger values of $x_0/x_s$, and is much faster than SMG.

\section{Elastic Constants \& Zone-Edge Phonons}

The elastic response of bcc and fcc lattices is obtained using small homogeneous strains $e_{ij}$.  In neutron star conditions, one cannot neglect hydrostatic pressure in a calculation of elastic constants $c_{ijkl}$ if one hopes to obtain accurate wave propagation speeds.  According to the standard work of \citet{bar65}
\begin{equation}
c_{ijkl} = \frac{1}{V_0}\frac{\partial^2 U}{\partial e_{ij}\partial e_{kl}} + \frac{P}{2}(2\delta_{ij}\delta_{kl} - \delta_{il}\delta_{jk} - \delta_{ik}\delta_{jl}), \label{cijkl}
\end{equation}
where $V_0$ is the volume of the reference state subject only to hydrostatic pressure and no other strains, and the strain derivative is taken with respect to this reference state.  In practice, one choses a strain matrix that isolates the elastic response to a particular $c_{ijkl}$, or a specific combination of them, and there are many ways to do this. Although one popular method \citep{ste04, gri12} employs certain volume-conserving strains such that the pressure correction term in Equation \ref{cijkl} vanishes, we use an alternative approach that keeps the numerical method as simple as possible. First we calculate $c_{11}=c_{xxxx}=c_{yyyy}=c_{zzzz}$ from a uniaxial strain.  The only modification to the numerical method is insertion of one or more extra layers of gridpoints at a height well away from any curved boundaries.  Next we calculate $c_{44}=c_{xyxy}=c_{yzyz}=c_{zxzx}$ using a symmetric shear deformation, i.e. the only nonzero strain components are $e_{yz}=e_{zy}=e_4/2$.  This method is equivalent to that given by Equations 2-3 in \citet{ste04}.  Shear deformation means the stencil, quadrature rule, domain and boundary conditions must all be modified, and owing to these complications we calculate $c_{44}$ for bcc only.  Our modified 7-point stencil has discretization in the shear plane accurate to order $h^2(1+(e_4/2)^2)$, and we are concerned only with values of $e_4\ll1$.  Symmetry lowering makes it necessary to increase the size of the domain to half of the conventional cell, and replace Neumann with periodic boundaries on surfaces normal to the shear plane.  \textit{Hypre}'s SMG solver has a power-of-two restriction on grid periods, so the sheared problems are computed using $1024^2\times513$ gridpoints.  The remaining independent elastic constant $c_{12}=c_{xxyy}=c_{yyzz}=c_{zzxx}$ is obtained from the single-crystal bulk modulus $K = -V\partial P/\partial V = (c_{11}+2c_{12})/3$. Cubic lattice stability requires that the Born criteria are met: $c'=c_{11}-c_{12}$, $c_{44}$, and $K$ must all be positive.

Elastic constants obtained in this manner contain no information about the Lorentz force acting on nuclei during lattice vibrations. This can drastically change the phonon spectrum of a Coulomb crystal -- see \citet{hae07} section 4.1.6b for a review, and more recently, \citet{bai09}.  At a semi-classical level we can expect the MTF solid to be less affected by this than the Coulomb crystal, because the nuclei are strongly screened by charge-neutralizing electrons. The effect may still be significant, however. With this caveat, we regard our prior calculation of elastic constants as a $k=0$ frozen phonon calculation, being, at present, less interested in the direct effect of the magnetic field on lattice vibrations than the indirect effect of the modified electron statistics. Other studies \citep{per06,chu07} have similarly neglected the distortion of the phonon spectrum by the magnetic field. \citet{chu07} have argued that this is justified at all but the strongest fields and lowest densities ($B\gtrsim10^{14}$ gauss and $\rho\lesssim10^6$ g/cc). 

Characteristic phase velocities of long wavelength phonons are given by 
\begin{eqnarray}
v_l & = & \sqrt{\frac{K}{\rho}}, \\
v_t & = & \sqrt{\frac{\mu_{\mathrm{eff}}}{\rho}},
\end{eqnarray}
for longitudinal and transverse modes, respectively.  In analysis of neutron star material, the effective shear modulus $\mu_{\mathrm{eff}}$ is typically taken to be the angle-averaged quantity proposed by \citet{oga90} -- see for example \citet{bai12, joh12}.  For cubic crystal symmetry, this takes the form
\begin{equation}
\mu_{\mathrm{eff}} = \frac{1}{5}(c_{11}-c_{12}+3c_{44}).
\end{equation}
Debye frequencies for longitudinal and transverse phonons can now be defined as
\begin{eqnarray}
\omega_{D,l} & = & k_D v_l, \label{omegaDl}\\
\omega_{D,t} & = & k_D v_t, \label{omegaDt}
\end{eqnarray}
where $k_D=(6\pi^2n_i)^{1/3}$ is the Debye wavenumber and $n_i$ is the number density of nuclei.  These quantities are not meant to suggest that the Debye model gives a good approximation to the actual phonon dispersion throughout the Brillouin zone, or that $\omega_D$ is an accurate zone-edge frequency.  Rather, they are intended merely as a characterization of the $k=0$ phonon dispersion.

To address the question of zone edge phonon frequencies in the MTF model, we perform a separate frozen phonon calculation at the H-point of the bcc Brillouin zone, where longitudinal and transverse branches coincide.  To visualize the frozen H-mode, start with the bcc conventional cell and slightly displace the nucleus in the body-center position towards a face-center position.  The domain and boundary conditions used for the $c_{44}$ calculation are conveniently recycled to treat this case (setting $e_4=0$).  However, the Ewald sum given by Equation \ref{latsum}, and associated Madelung constant, must be generalized to a non-Bravais lattice, as in \citet{dar80}.  The squared H-mode frequency $\omega_H^2$ is given by the ratio of the frozen phonon spring constant to the reduced mass of the system of interpenetrating simple cubic sublattices.  Again, the effect of Lorentz forces on nuclei is ignored.

\section{Results}

\begin{figure}[b]
\plotone{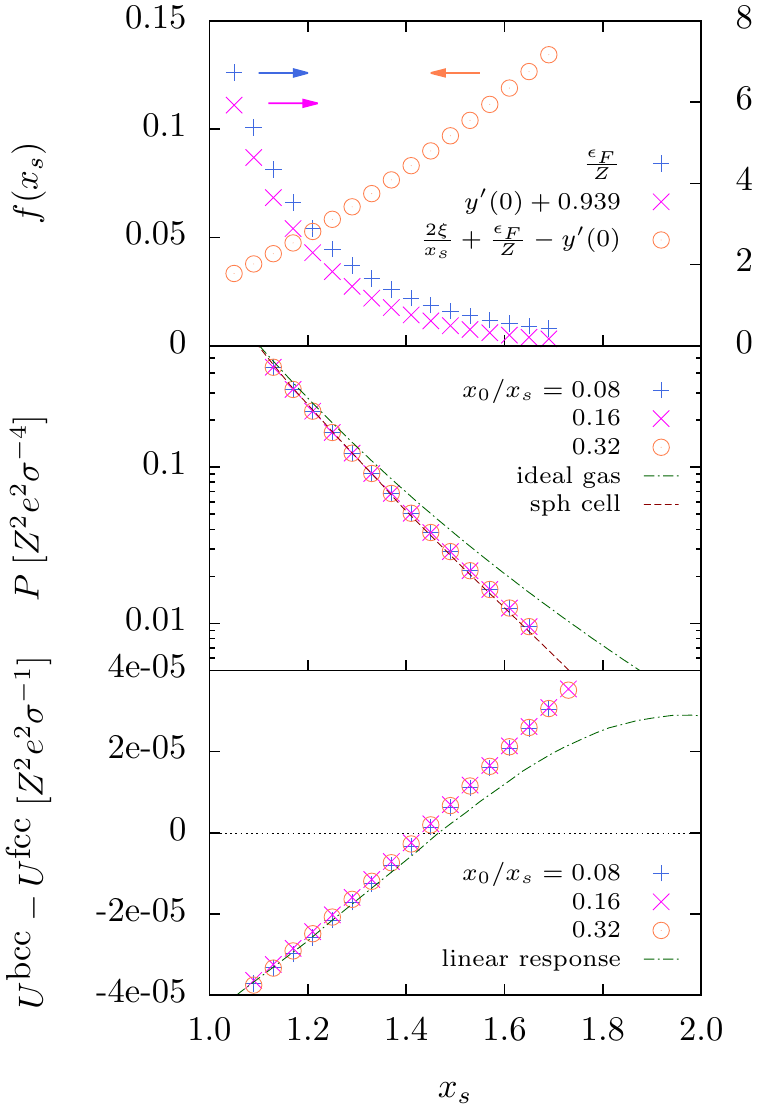}
\caption{Universal MTF solution for the bcc lattice (unless fcc is specified). Top: Behavior of $\epsilon_F/Z$, the derivative condition, and the nonuniformity correction, computed with $x_0/x_s=0.08$. The values $\epsilon_F=0$ and $y'(0)=-0.939$ correspond to the MTF ``atom" \citep{ban74}.  Middle: equation of state.  The dash-dot line is the lowest Landau level ideal gas equation of state, and the dashed line is the MTF equation of state in the spherical Wigner-Seitz cell approximation \citep{fus89}.  Bottom: $T=0$ Helmholtz free energy difference between bcc and fcc lattice, per nucleus.  The dash-dot line shows the linear response theory result for lowest Landau level occupation, see Equation \ref{lrt} and  \citet{bai02}, for example.
\label{fig:mtfuni}}
\end{figure}

For comparison with the non-periodized model and linear response theory, calculations are performed with $x_s$ in the range 1 to 2. An issue arises for $x_s > 1.73$, where an intermediate stage of Newton's iteration generates an approximate solution $u_k$ that is not everywhere real.  No results are presented in these cases.  We first check that the asymptotic limits of $\epsilon_F/Z$ and $y'(0)$ are consistent with the MTF ``atom" at zero pressure, and that the nonuniformity correction $2\xi/x_s + \epsilon_F/Z - y'(0)$ tends to zero in the high density limit, see Figure \ref{fig:mtfuni}.  
Next we compute the equation of state and $T=0$ free energy difference between bcc and fcc structures, also shown in Figure \ref{fig:mtfuni}.  The equation of state is very close to that obtained using spherical Wigner-Seitz cells, with a slight hardening at low density.  At high density, good agreement is also found with the linear response bcc-fcc energy difference, whereas at low density there is a notable departure towards favoring fcc.  This is the first indication that the equilibrium phase diagram may be significantly different than that predicted by linear response of the lowest Landau level electron gas.  Surprisingly, we find almost no dependence of these results on the Dirichlet radius $x_0$, over two doublings.  (The bcc-fcc energy difference in particular should be a good test of $x_0$-dependence). We can thus be reasonably certain that the data shown represent the $x_0\to0$ limit.

\begin{figure}[t]
\plotone{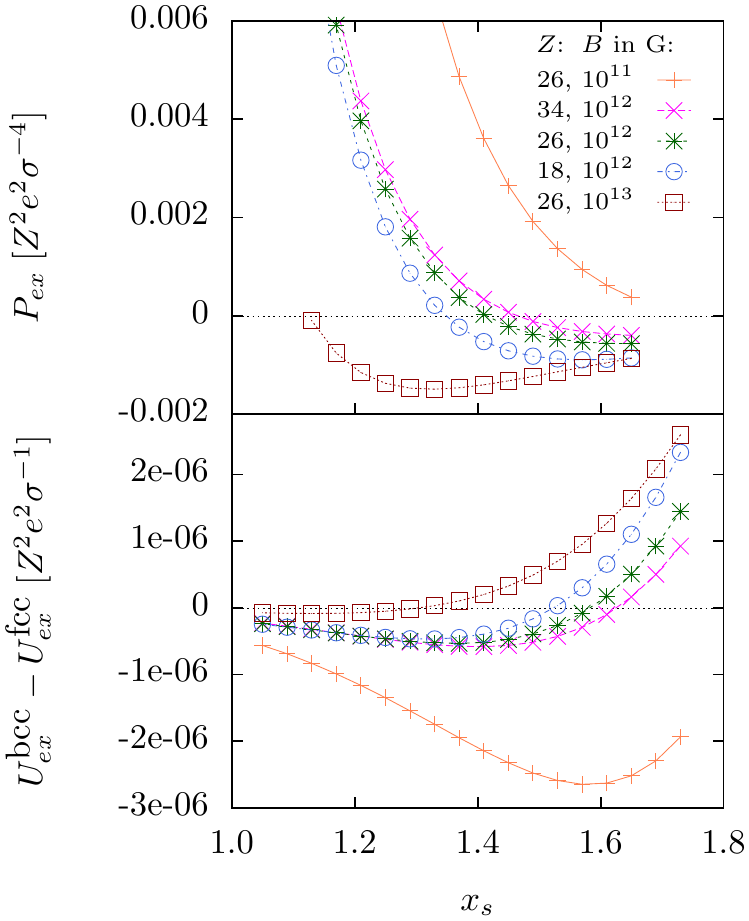}
\caption{Exchange correction to the MTF model computed with $x_0/x_s=0.16$.  The exchange functional used is the leading order term obtained by \citet{dan71} and \citet{fus89}. Top: correction to the bcc lattice equation of state for several values of nuclear charge $Z$ and field strength $B$ (universality is lost with the inclusion of exchange).  Bottom: correction to $T=0$ Helmholtz free energy difference between bcc and fcc lattice, per nucleus.  Symbols are the same as above.
\label{fig:mtfd}}
\end{figure}

Next we calculate an exchange correction to the equation of state and bcc-fcc energy difference to zeroth order in $\delta n_e = n_e^{TFD}-n_e^{TF}$, where $n_e^{TFD}$ and $n_e^{TF}$ are the self-consistent Thomas-Fermi-Dirac and Thomas-Fermi densities.  A local density approximation for the exchange energy of a strongly magnetized electron gas was first given by \citet{dan71}.  To leading order in the low density expansion, their result agrees with that later obtained by \citet{fus89}  
\begin{equation}
E_{ex} = \frac{b^2}{2\pi^3}\frac{e^2}{a_0^4} \int d^3r \;\Big(\frac{n_e}{n_*}\Big)^2 \Big[2\ln\Big(\frac{2n_e}{n_*}\Big) + \gamma - 3\Big]. \label{exchange}
\end{equation}
Here $\gamma=0.5772\dots$ is the Euler constant and $n_* = 2^{-1/2}\pi^{-2}\ell^{-3}$ is the density at which the first excited Landau level starts to become populated.  For the purpose of our crude zeroth-order calculation, this leading order term is sufficient and avoids the non-integrable divergence at the nuclei that one would get with higher order terms.  Results for the exchange correction are shown in Figure \ref{fig:mtfd}.  For high field strengths and low densities, the exchange correction slightly softens the equation of state and further increases the energetic favorability of fcc.  

\begin{figure}[t]
\plotone{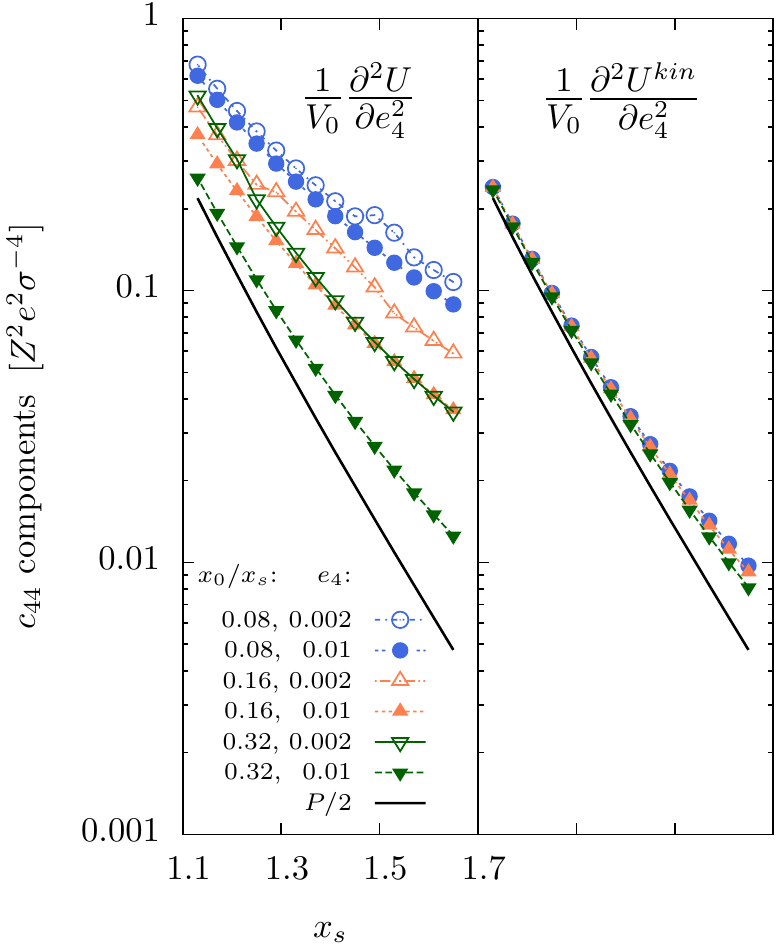}
\caption{  
Left panel: Dependence of $c_{44}$ on the Dirichlet radius (symbol color) and strain magnitude (open vs filled symbols).  The pressure correction $-P/2$, coming from the last term of Equation \ref{cijkl}, has been subtracted out and is shown separately by the black line.  Evidently the $c_{44}$ calculation has not fully converged, although the values obtained at different strain magnitudes clearly become more consistent as $x_0\to0$.  Right panel: Electron kinetic energy contribution to the strain derivative, i.e. to the first term in Equation \ref{cijkl}.  Symbols have the same meaning as in the left panel.  At high densities, the pressure correction largely cancels the kinetic contribution to $c_{44}$. \label{fig:c44}}
\end{figure}

\begin{figure}[t]
\plotone{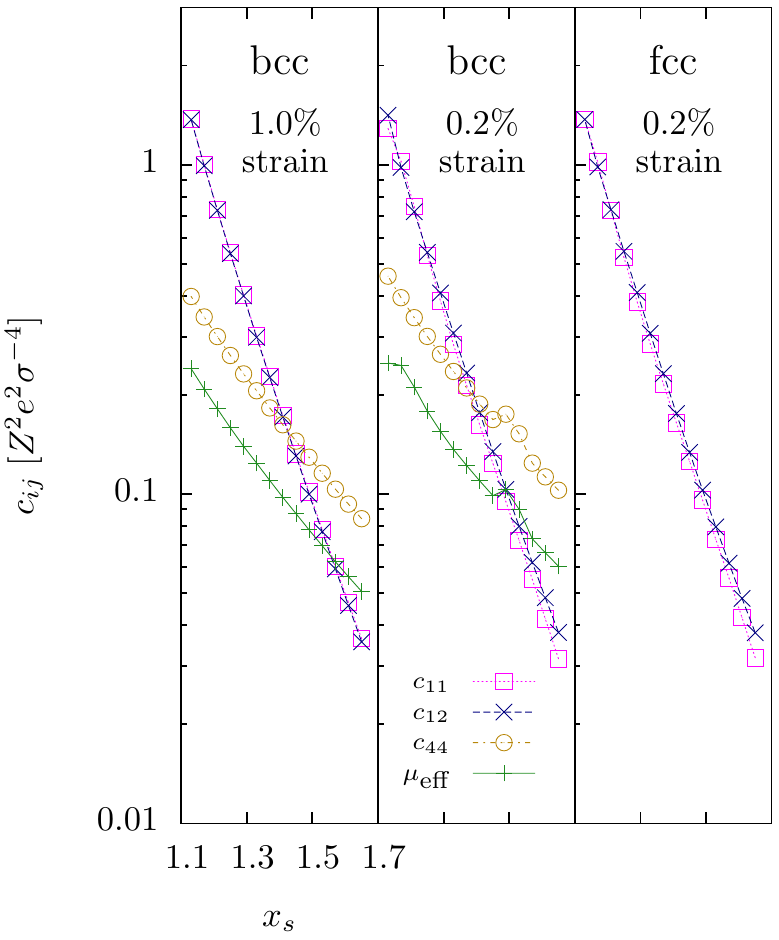}
\caption{  
Elastic constants for bcc (left \& center panels) and fcc (right panel) using the smallest Dirichlet radius ($x_0/x_s=0.08$).  In the case of uniaxial strain, the small strain magnitude ($\approx0.2$\%) corresponds to stretching and compressing the domain by one grid spacing.  These small strain data suggest that both bcc and fcc are unstable (or at best marginally stable) against a symmetry-lowering transition to an orthorhombic or tetragonal structure, due to $c'=c_{11}-c_{12}<0$, except perhaps at the highest densities. No $c_{44}<0$ or $K<0$ instabilities were found. \label{fig:elastic}}
\end{figure}

Elastic constants are more sensitive to the Dirichlet radius than are the static quantities so far considered.  They also show some dependence on the strain magnitude.  In the case of $c_{11}$ and $c_{12}$, these combined dependences are weak, amounting to variations on the order of $10\%$ over the range of Dirichlet radius and strain magnitude studied.  In the case of $c_{44}$, the dependencies are stronger (see Figure \ref{fig:c44}).  We reiterate that $x_0\to0$ is the desired limit for the 3D MTF model.  Fortunately, $c_{44}$'s spurious strain dependence tends to go away as this limit is approached and reasonably consistent results (across different strain magnitudes) are obtained at the smallest Dirichlet radius ($x_0/x_s=0.08$).  Figure \ref{fig:c44} also shows that the electron kinetic contribution to $c_{44}$ is in good agreement with the magnitude of the pressure correction $-P/2$, at high density.  More will be said about this at the end of the section.  Our best-converged results for the $c_{ij}$, obtained at $x_0/x_s=0.08$, are shown in Figure \ref{fig:elastic}, and the corresponding elastic moduli given in Table \ref{tab:moduli}.  In Table \ref{tab:moduli} and for the remaining analysis, we take $\mu_{\mathrm{eff}}$ as the smoother of the two data series obtained at $x_0/x_s=0.08$ (green crosses in the left panel of Figure \ref{fig:elastic}).  

Remarkably similar elastic constants are found for bcc and fcc lattices.  Both lattices exhibit instabilities due to $c'=c_{11}-c_{12}<0$. Since $c'$ describes the response to a deformation involving a diagonal, zero-trace strain matrix, this suggests that the true equilibrium structure is either orthorhombic or tetragonal. This type of lattice instability also occurs in terrestrial metals: both cubic phases of the light actinides uranium and neptunium suffer from $c'<0$ and adopt orthorhombic structures in equilibrium \citep{gri12}.  We leave the search for the stable structure, i.e. mapping out Bain transformation paths, for future work. If the cubic lattices are instead marginally stable, as the results at larger strain and/or larger $x_0$ suggest, they appear to be highly anisotropic in the sense $c'/2c_{44}\ll1$.  This is consistent with the Coulomb crystal, which is elastically similar to lithium and plutonium \citep{kob14}.  No $K<0$ or $c_{44}<0$ instabilities were found at any $x_0$, although the latter criterion has been checked only for bcc. A consideration of soft modes associated with potential incipient lattice instabilities is interesting but beyond the scope of the present work.

\begin{figure}[h]
\plotone{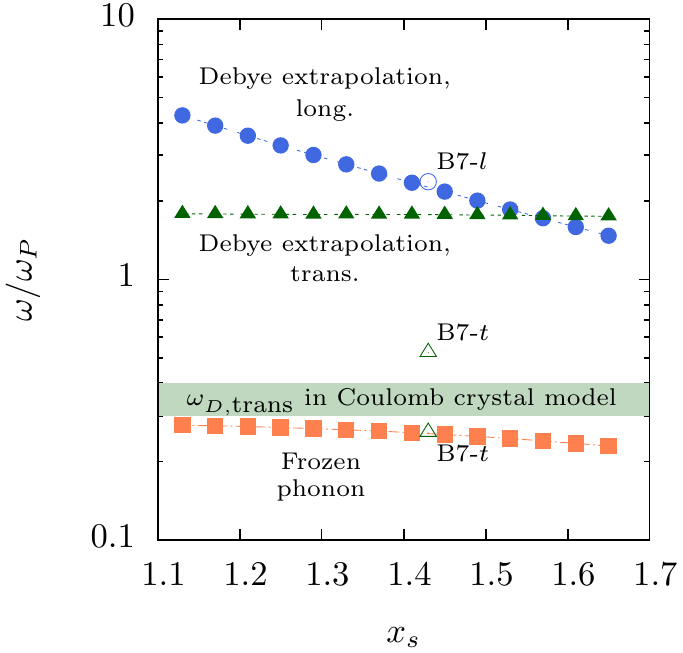}
\caption{  
A linear (Debye) extrapolation of the $k=0$ acoustic modes to the bcc Brillouin zone edge overestimates the zone-edge frequencies calculated directly by a frozen phonon method.  Filled blue circles, green triangles, and orange squares correspond to $\omega_{D,l}$, $\omega_{D,t}$, and $\omega_H$, respectively.  Data points marked with open symbols are extracted from \citet{bai02} Figure 7, which shows the fcc phonon dispersion in the linear response approximation, in a low symmetry direction and for $k_{TF}r_s=1$.  The shaded green band indicates the range of transverse Debye frequencies appropriate to Coulomb crystal models -- see for example, \citet{chu07, cha92}.\label{fig:thermal}}
\end{figure}

\begin{table}[ht]
\caption{
Pressure, bulk modulus, and effective shear modulus of the bcc MTF solid, in units of $Z^2e^2/\sigma^4$.
}
\centering
\begin{tabular}{c c c c}
\hline\hline
$x_s$ & $P$ & $K$ & $\mu_{\mathrm{eff}}$ \\ [0.5ex]
\hline
1.13 & 0.437 & 1.372 & 0.240 \\
1.17 & 0.315 & 0.995 & 0.208 \\
1.21 & 0.228 & 0.729 & 0.182 \\
1.25 & 0.167 & 0.538 & 0.159 \\
1.29 & 0.123 & 0.401 & 0.140 \\
1.33 & 0.0912 & 0.300 & 0.124 \\
1.37 & 0.0679 & 0.227 & 0.110 \\
1.41 & 0.0509 & 0.172 & 0.0978 \\
1.45 & 0.0382 & 0.131 & 0.0872 \\
1.49 & 0.0288 & 0.100 & 0.0779 \\
1.53 & 0.0218 & 0.0773 & 0.0697 \\
1.57 & 0.0165 & 0.0597 & 0.0623 \\
1.61 & 0.0125 & 0.0462 & 0.0561 \\
1.65 & 0.00954 & 0.0359 & 0.0506 \\ [0.5ex]
\hline
\end{tabular}
\label{tab:moduli}
\end{table}
Longitudinal and transverse Debye frequencies (extrapolated from the Brillouin zone center) and H-mode frequency $\omega_H$ are given in Figure \ref{fig:thermal} in units of the nuclear (ion) plasma frequency $\omega_P=\sqrt{4\pi n_i Z^2e^2/M}$.  As with the static lattice properties, $\omega_H$ was found to have negligible $x_0$-dependence.  Consistent results were also found across different frozen phonon amplitudes.  The picture we have thus obtained of the transverse phonon dispersion in the nonlinear MTF model is the following:  At $k=0$, the dispersion is apparently 5-7 times stronger than in the unscreened Coulomb crystal model (compare the green triangles with the shaded green band in Figure \ref{fig:thermal}).  This $v_t$ enhancement seemingly cannot be explained by linear response screening corrections to the Coulomb crystal model, because transverse phonons are largely insensitive to linear response screening, and the small correction obtained tends to decrease rather than increase $v_t$ \citep{bai02, bai12}.  At the zone edge, however, the transverse phonon frequencies come in rather close to the Coulomb crystal's Debye frequency (compare the orange squares with the shaded green band in Figure \ref{fig:thermal}).  Since we have only considered the zone center and zone edge, the wavenumber $k'$ at which the spectrum begins to deviate strongly from a linear dispersion is not known.  An upper limit is $k'\approx k_D/2$, since if the linear dispersion held farther from the center of the Brillouin zone (Debye sphere), phonon frequencies would exceed $\omega_P$.  Answering this question within the framework of the iMH method would require inclusion of a large number of unit cells in the computational domain.

Because our calculation of $c_{44}$ involves a volume change and the strongly magnetized, degenerate electron gas is much less compressible than the nonmagnetized gas (due to its higher adiabatic index), it is important to confirm that the $v_t$ enhancement in the MTF solid is not an artifact of our particular choice of strain deformation. The Helmholtz energy (Equation \ref{helmholtz}) can be decomposed into kinetic and electrostatic terms, so we similarly decompose $c_{44}=c_{44}^{kin}+c_{44}^{es}+c_{44}^{pc}$ (where the pressure correction $c_{44}^{pc}$ is given by the last term in Equation \ref{cijkl}).  Components $c_{44}^{kin}$ and $c_{44}^{pc}$ are both nonzero due to the volume change $V_0(e_4/2)^2$ accompanying our shear deformation; in contrast, they both vanish for the volume-conserving shear deformation given by Equation 4 in \citet{ste04}. Figure \ref{fig:mtfuni} also shows that the total pressure $P$ entering into $c_{44}^{pc}$ is, at high densities, dominated by the electron kinetic pressure. These observations suggest a correspondence between $c_{44}^{kin}$ and $c_{44}^{pc}$.  Indeed, we find that these two quantities cancel almost completely at high densities (see the right panel of Figure \ref{fig:c44}). In addition, $c_{44}^{kin}$ is several times smaller than $c_{44}^{es}$ in the appropriate limit of small Dirichlet radius. Thus we conclude that $c_{44}\approx c_{44}^{es}$ regardless of which strain deformation is used. We also would not expect $c_{44}^{es}$ to be significantly different in the volume-conserving method, since the only extra strain component involved in that case is to leading order the square of the symmetric shear strain component, meaning that for small strains the size and shape of the unit cell is very similar in the two methods.

\section{Discussion}

The comparable magnitudes of $v_l$ and $v_t$ in the MTF model may be unusual in the context of polarizable plasmas, but there are many examples of atomic solids where a similar situation occurs, and the MTF model is, after all, a crude description of an atomic solid in a huge magnetic field.  Enhancement of $v_t$ by a factor of 5-7 could have consequences for magnetized neutron star envelopes at $T\ll T_P/3$, where $T_P=\hbar\omega_P/k_B$ is the ion plasma temperature.  In this temperature regime, only the lowest-frequency acoustic phonons around $k=0$ are thermally occupied, and those of the MTF solid are substantially stiffer than those of the Coulomb crystal. The sound speed is therefore much higher, while the number of thermally excited phonon modes available for thermal transport at a given temperature is much lower.

The elastic instabilities we have found support mounting evidence that the crystal lattice structure of a neutron star crust is more complicated than heretofore assumed, across a range of depths and associated physical regimes \citep{kob14}.  Even the simple MTF model, which contains no explicit symmetry-breaking mechanism, can result in an equilibrium structure with unexpectedly low symmetry.  Aside from obvious thermodynamic signatures such as latent heat, a low-symmetry structure can couple to the magnetic field direction. If the coupling is strong, one can imagine the transition being driven by a changing field, or conversely, the transition exerting a back-action on the crustal field, which could have non-local effects.  Also, a $c'<0$ driven transition could potentially be of a martensitic (shape memory) nature.

Finally, we suggest that the domain decomposition method described here could be extended to treat certain ``pasta phases" in neutron star cores, complementary to dimensional continuation techniques \citep{joh12}.

\acknowledgments
This work used the Extreme Science and Engineering Discovery Environment (XSEDE), which is supported by National Science Foundation grant number ACI-1053575.  T.A.E. acknowledges an Academic Computing Fellowship from The Pennsylvania State University.  We thank George Pavlov and an anonymous referee for useful comments.

\end{document}